\documentclass[12pt]{article}
\usepackage{latexsym,amssymb,amsfonts,amsmath}
\usepackage[dvips]{graphicx}
\usepackage[usenames]{color}
\usepackage{dsfont}
\usepackage[english]{babel}
\usepackage{graphicx}
\usepackage{wrapfig}
\usepackage{epstopdf}
\usepackage{subfig}
\usepackage{fancyhdr}
\usepackage{mcite}
\usepackage{amsthm}
\usepackage{eso-pic}

\newlength{\extraspace}
\newlength{\extraspaces}
\newcommand{\be}{\begin{equation}
\addtolength{\abovedisplayskip}{\extraspaces}
\addtolength{\belowdisplayskip}{\extraspaces}
\addtolength{\abovedisplayshortskip}{\extraspace}
\addtolength{\belowdisplayshortskip}{\extraspace}}
\newcommand{\ee}{\end{equation}}

\newcommand{\bea}{\begin{eqnarray}
\addtolength{\abovedisplayskip}{\extraspaces}
\addtolength{\belowdisplayskip}{\extraspaces}
\addtolength{\abovedisplayshortskip}{\extraspace}
\addtolength{\belowdisplayshortskip}{\extraspace}}
\newcommand{\eea}{\end{eqnarray}}

\newcommand{\newsection}[1]{
\pagebreak[3]
\addtocounter{section}{1}
\setcounter{equation}{0}
\setcounter{subsection}{0}
\setcounter{footnote}{0}
\begin{flushleft}
{\large\bf \thesection. #1}
\end{flushleft}
\nopagebreak
\nopagebreak}

\newcommand{\newsubsection}[1]{
\vspace{5mm}
\pagebreak[3]
\addtocounter{subsection}{1}
\noindent{\bf \thesubsection. #1}
\nopagebreak
\vspace{2mm}
\nopagebreak}

\newcommand{\newsubsubsection}[1]{
\vspace{5mm}
\pagebreak[3]
\addtocounter{subsubsection}{1}
\noindent{\bf \thesubsubsection. #1}
\nopagebreak
\vspace{2mm}
\nopagebreak}

\def\Tr{{\rm Tr}}

\definecolor{mygray}{gray}{0.5}
\definecolor{mygray2}{gray}{0.3}

\newcommand{\virg}[1]{``#1''}
\newcommand{\trpl}[1]{\vec{#1}_\perp}

\newcommand{\mwf}[1]{|\psi_{#1}(\trpl{R_{#1}},f_{#1})|^2}
\newcommand{\wfint}{\int d^2 \trpl{R_1}\,\int_{0}^{1}df_1\, \mwf{1} \,\int d^2 \trpl{R_2}\,\int_{0}^{1}df_2\, \mwf{2}}
\newcommand{\mvar}[1]{\trpl{R_{#1}},f_{#1}}
\renewcommand{\to}{\rightarrow}
\newcommand{\re}{\text{Re}}
\newcommand{\im}{\text{Im}}

\begin{document}

\addtolength{\baselineskip}{.8mm}

{\thispagestyle{empty}

%\noindent \hspace{1cm} \hfill IFUP--TH/2012--03 \hspace{1cm}\\
%\mbox{}                \hfill September 2012 \hspace{1cm}\\
%\mbox{}                \hfill Expanded version \hspace{1cm}\\

\begin{center}
\vspace*{1.0cm}
{\large\bf ASYMPTOTIC ENERGY DEPENDENCE OF HADRONIC TOTAL CROSS SECTIONS
FROM LATTICE QCD}
\\
\vspace*{1.0cm}
{\large Matteo Giordano$^{1,}$\footnote{E--mail: giordano@unizar.es},
Enrico Meggiolaro$^{2,}$\footnote{E--mail: enrico.meggiolaro@df.unipi.it},
Niccol\`o Moretti$^{2}$\footnote{E--mail: nicco.moretti@gmail.com}
}\\
\vspace*{0.5cm}{\normalsize
{$^{1}$ Departamento de F\'\i sica Te\'orica, Universidad de Zaragoza,\\
Calle Pedro Cerbuna 12, E--50009 Zaragoza, Spain}}\\
\vspace*{0.5cm}{\normalsize
{$^{2}$ Dipartimento di Fisica, Universit\`a di Pisa,
and INFN, Sezione di Pisa,\\
Largo Pontecorvo 3, I--56127 Pisa, Italy}}\\
\vspace*{2cm}{\large \bf Abstract}
\end{center}

\noindent
The nonperturbative approach to {\it soft} high--energy hadron--hadron
scattering, based on the analytic continuation of Wilson--loop correlation
functions from Euclidean to Minkowskian theory, allows to investigate
the asymptotic energy dependence of hadron--hadron total cross sections
in lattice QCD. In this paper we will show, using best fits of the lattice
data with proper functional forms satisfying unitarity and other physical
constraints, how indications emerge in favor of a universal asymptotic
high--energy behavior of the kind $B \log^2 s$ for hadronic total
cross sections.
\\
}
\newpage

\newsection{Introduction}

The problem of predicting total cross sections
at high energy from first principles is one of the oldest open
problems of hadronic physics, not yet satisfactorily solved in QCD.
Present--day experimental observations (up to a center--of--mass total energy
$\sqrt{s} = 7$ TeV, reached at the LHC $pp$ collider \cite{TOTEM}) seem to
support the following asymptotic high--energy behavior:
$\sigma_{\rm tot}^{(hh)} (s) \sim B \log^2 s$,
with a {\it universal} (i.e., {\it not} depending on the particular hadrons
involved) coefficient $B \simeq 0.3$ mb \cite{Blogs}.
This behavior is consistent with the
well--known {\it Froissart--Lukaszuk--Martin} (FLM) {\it theorem}~\cite{FLM},
according to which, for $s \to \infty$,
$\sigma^{(hh)}_{\rm tot}(s) \le {\frac{\pi}{m_\pi^2}}
\log^2 \left( {\frac{s}{s_0}} \right)$,
where $m_\pi$ is the pion mass and $s_0$ is an unspecified squared mass scale.
(Let us observe that the {\it experimental} value of $B$ is much smaller
than [about $0.5 \%$] the coefficient ${\frac{\pi}{m_\pi^2}}$ appearing
in the FLM bound.)
As we believe QCD to be the fundamental theory of strong interactions,
we also expect that it correctly predicts from first principles the behavior
of hadronic total cross sections with energy. Anyway, in spite of all the
efforts, a satisfactory solution to this problem is still lacking.
Theoretical supports to the universality of the coefficient $B$ were
found in the model of the iteration of {\it soft--Pomeron} exchanges by
eikonal unitarization \cite{soft-pomeron} (recently revisited in the
context of holographic QCD \cite{BKYZ}), and also using arguments
based on the so--called {\it Color Glass Condensate} of QCD \cite{CGC},
or simply modifying the original {\it Heisenberg's model} \cite{Heisenberg}
in connection with the presence of glueballs \cite{DGN}.

This problem is part of the more general problem of high--energy elastic
scattering at low transferred momentum, the so--called {\it soft
high--energy scattering}. As soft high--energy processes possess two
different energy scales, the total center--of--mass energy squared
$s$ and the transferred momentum squared $t$, smaller than the typical
energy scale of strong interactions ($|t| \lesssim 1~ {\rm GeV}^2 \ll
s$), we cannot fully rely on perturbation theory. A genuine
nonperturbative approach in the framework of QCD has been proposed
in~\cite{Nachtmann91} and further developed in a number of papers (see, e.g.,
\cite{pomeron-book} for a review and a complete list of references): 
using a functional integral approach, high--energy hadron--hadron
elastic scattering amplitudes are shown to be governed by the
correlation function (CF) of certain Wilson loops defined in Minkowski space
\cite{DFK,Nachtmann97,BN,Dosch,LLCM1}.
Moreover, as it has been shown in \cite{analytic,Meggiolaro05,GM2009},
such a CF can be reconstructed by \emph{analytic continuation} from the CF
of two Euclidean Wilson loops, that can be calculated using the nonperturbative
methods of Euclidean Field Theory. 

The analytic--continuation relations have allowed the nonperturbative
investigation of correlators (and the corresponding scattering amplitudes)
using some analytical models, such as the \emph{Stochastic Vacuum Model}
(SVM) \cite{LLCM2}, the \emph{Instanton Liquid Model} (ILM)
\cite{ILM,GM2010}, the AdS/CFT \emph{correspondence} \cite{JP} and, finally,
they have also allowed a numerical study by Monte Carlo simulations in
\emph{Lattice Gauge Theory} (LGT) \cite{GM2008,GM2010} (see also Refs.
\cite{GM2011} for a short review).
Although the numerical results obtained on the lattice can
be considered \virg{exact} (since they are derived from \emph{first principles}
of QCD), it is not possible to relate them directly to physical quantities,
since the analytic continuation of the correlator can be performed only if an
analytical dependence on the variables is known, while lattice data can be
obtained only for a discrete (finite) set of values. However, it is possible
to test the goodness of the known analytical models (like the ones we have
mentioned above) simply through a best fit to the lattice data. This analysis
has already been done in Refs. \cite{GM2008,GM2010} and it is briefly recalled
in Section 3. The result of this analysis is not, generally speaking,
satisfactory: known analytical models lead to bad quality best fits and,
moreover, none of them provides a physically acceptable total cross section.
%(SVM and ILM lead to constant total cross sections at high energy, while the
%AdS/CFT correspondence leads to a power--like behavior
%$\sigma_{\text{tot}}\sim s^{\frac{1}{3}}$ that explicitly violates the
%FLM bound, which is expected to be valid for the hadronic scattering.)

In this paper, after a brief survey (for the benefit of the reader) of the
nonperturbative approach to soft high--energy scattering in the case of
meson--meson {\it elastic} scattering (in Section 2), and of
the numerical approach based on LGT, comparing the numerical results
to the existing analytical models (in Section 3), we will concentrate
on the search for a new parameterization of the
(Euclidean) correlator that, in order: $i)$ fits well the lattice data; $ii)$
satisfies (after analytic continuation) the unitarity condition; and, most
importantly, $iii)$ leads to a rising behavior of total cross sections at
high energy, in agreement with experimental data. In particular, one is
interested in the dependence of the CF on the angle $\theta$
between the loops, since it is related, after analytic continuation, to the
energy dependence of the scattering amplitudes, and also in its dependence on
the impact--parameter distance. In Section 4 we show that, making some
reasonable assumptions about the angular dependence and the impact--parameter
dependence of the various terms in the parameterization, our approach leads
quite \virg{naturally} to total cross sections rising asymptotically as
$B \log^2 s$ (that is what experimental data seem to suggest). Moreover,
in our approach the coefficient $B$ turns out to be \emph{universal}, i.e,
the same for all hadronic scattering processes (as it also seems to be
suggested by experimental data), being related to the mass--scale $\mu$ which
sets the large impact--parameter behavior of the correlator.
%(and which is expected to be proportional to the lightest glueball mass $M_G$
%or to the inverse of the so--called \virg{\emph{vacuum correlation length}}
%$\lambda_{vac}$ \cite{lambda-vacuum,DDSS}).
This is actually the main result of this paper.
In Section 5 we draw our conclusions and discuss some prospects for the future.

\newpage

\newsection{High--energy meson--meson elastic scattering amplitude
and Wilson--loop correlation functions} 

We sketch here the nonperturbative approach to soft high--energy
scattering (see~\cite{GM2008} for a more detailed presentation). The
elastic scattering amplitude ${\cal M}_{(hh)}$ of two hadrons, or more
precisely {\it mesons} (taken for simplicity with the same mass $m$),
in the {\it soft} high--energy regime can be reconstructed, after folding
with two proper squared hadron wave functions $|\psi_1|^2$ and $|\psi_2|^2$,
describing the two interacting hadrons, from the scattering amplitude
${\cal M}_{(dd)}$ of two dipoles of fixed transverse sizes
$\vec{R}_{1,2\perp}$, and fixed longitudinal--momentum fractions $f_{1,2}$ of
the two quarks in the two dipoles \cite{DFK,Nachtmann97,BN,Dosch,LLCM1}:
\bea
{\cal M}_{(hh)}(s,t) &=&
\displaystyle\int d^2\vec{R}_{1\perp} \int_0^1 df_1~
|\psi_1(\vec{R}_{1\perp},f_1)|^2
\displaystyle\int d^2\vec{R}_{2\perp} \int_0^1 df_2~
|\psi_2(\vec{R}_{2\perp},f_2)|^2
\nonumber \\
&\times& {\cal M}_{(dd)} (s,t;\vec{R}_{1\perp},f_1,\vec{R}_{2\perp},f_2) ,
\label{scatt-hadron}
\eea
with: $\int d^2\vec{R}_{1\perp} \int_0^1 df_1~
|\psi_1(\vec{R}_{1\perp},f_1)|^2 =
\int d^2\vec{R}_{2\perp} \int_0^1 df_2~
|\psi_2(\vec{R}_{2\perp},f_2)|^2 = 1$.\\
(For the treatment of baryons, a similar, but more involved,
picture can be adopted, using a genuine three--body configuration or,
alternatively and even more simply, a quark--diquark configuration: we refer
the interested reader to the above--mentioned original
references~\cite{DFK,Nachtmann97,BN,Dosch,LLCM1}.) \\
In turn, the dipole--dipole ({\it dd}) scattering amplitude is obtained from
the (properly normalized) CF of two Wilson loops in the fundamental
representation, defined in Minkowski spacetime, running along the paths made
up of the quark and antiquark classical straight--line trajectories, and thus
forming a hyperbolic angle $\chi \simeq \log(s/m^2)$ in the longitudinal
plane (see Fig. 1). The paths are cut at proper times $\pm T$ as an infrared
regularization, and closed by straight--line ``links'' in the transverse plane,
in order to ensure gauge invariance. Eventually, the limit $T\to\infty$ has to
be taken. It has been shown in \cite{analytic,Meggiolaro05,GM2009}
that the relevant Minkowskian CF ${\cal G}_M(\chi;T;\vec{z}_\perp;1,2)$
($\vec{z}_\perp$ being the {\it impact parameter}, i.e., the transverse
separation between the two dipoles) can be reconstructed, by means of
{\it analytic continuation}, from the Euclidean CF of two Euclidean
Wilson loops, 
\begin{equation}
\label{GE}
{\cal G}_E(\theta;T;\vec{z}_\perp;1,2) \equiv
\dfrac{\langle \widetilde{\cal W}^{\,(T)}_1 \widetilde{\cal W}^{\,(T)}_2
\rangle_E}{\langle \widetilde{\cal W}^{\,(T)}_1 \rangle_E
\langle \widetilde{\cal W}^{\,(T)}_2 \rangle_E } - 1\,, \,\,\,\,
\widetilde{\cal W}^{\,(T)}_{1,2} \equiv 
{\displaystyle\dfrac{1}{N_c}} \Tr \left\{ {T}\!\exp
\left[ -ig \displaystyle\oint_{\widetilde{\cal C}_{1,2}}
 \tilde{A}_{\mu}(\tilde{x}) d\tilde{x}_{\mu} \right] \right\},
\end{equation}
where $\langle\ldots\rangle_E$ is the average in the sense of the
Euclidean QCD functional integral, and the arguments ``$1[2]$'' in
${\cal G}_E$ (and ${\cal G}_M$) stand for ``$\vec{R}_{1[2]\perp}, f_{1[2]}$''.
The Euclidean Wilson loops
$\widetilde{\cal W}^{\,(T)}_{1,2}$ are calculated on the following
quark $[q]$ -- antiquark $[\bar{q}]$ straight--line paths,
\begin{equation}
\widetilde{\cal C}_1 : \tilde{X}_1^{q[\bar{q}]}(\tau) = \tilde{z} +
\frac{\tilde{p}_{1}}{m} \tau 
+ f^{q[\bar{q}]}_1 \tilde{R}_{1} , \quad
\widetilde{\cal C}_2 : \tilde{X}_2^{q[\bar{q}]}(\tau) =
\frac{\tilde{p}_{2}}{m} \tau 
+ f^{q[\bar{q}]}_2 \tilde{R}_{2},
\label{trajE}
\end{equation}
with $\tau\in [-T,T]$, and closed by straight--line paths in the
transverse plane at $\tau=\pm T$. The four--vectors $\tilde{p}_{1}$
and $\tilde{p}_{2}$ are chosen to be $\tilde{p}_{1,2}={m}(
\pm\sin\frac{\theta}{2}, \vec{0}_{\perp}, \cos\frac{\theta}{2})$, $\theta$
being the angle formed by the two trajectories, i.e., $\tilde{p}_{1} \cdot
\tilde{p}_{2} = m^2 \cos\theta$. Moreover, $\tilde{R}_{i} =
(0,\vec{R}_{i\perp},0)$, $\tilde{z} = (0,\vec{z}_{\perp},0)$ and
$f_i^q \equiv 1-f_i$, $f_i^{\bar{q}} \equiv -f_i$.
We define also the Euclidean and Minkowskian CFs with the infrared cutoff
removed as
\bea
\displaystyle {\cal C}_E(\theta;\vec{z}_\perp;1,2) &\equiv& \lim_{T\to\infty}
{\cal G}_E(\theta;T;\vec{z}_\perp;1,2) ,
\nonumber \\
\displaystyle {\cal C}_M(\chi;\vec{z}_\perp;1,2) &\equiv& \lim_{T\to\infty}
{\cal G}_M(\chi;T;\vec{z}_\perp;1,2) .
\eea
The {\it dd} scattering amplitude is then obtained from
${\cal C}_E(\theta;\ldots)$ [with $\theta\in(0,\pi)$] by means
of analytic continuation as
\bea
\lefteqn{
{\cal M}_{(dd)} (s,t;1,2) 
\equiv -i~2s \displaystyle\int d^2 \vec{z}_\perp
e^{i \vec{q}_\perp \cdot \vec{z}_\perp}
{\cal C}_M(\chi \simeq \log(s/m^2); \vec{z}_\perp;1,2) }
\nonumber \\
& & = -i~2s \displaystyle\int d^2 \vec{z}_\perp
e^{i \vec{q}_\perp \cdot \vec{z}_\perp}
{\cal C}_E(\theta\to -i\chi \simeq -i\log(s/m^2); \vec{z}_\perp;1,2)\, ,
\label{scatt-loop}
\eea
where $s$ and $t = -|\vec{q}_\perp|^2$ ($\vec{q}_\perp$
being the transferred momentum) are the usual Mandelstam variables
(for a detailed discussion on the analytic continuation
see~\cite{GM2009}, where we have shown, on nonperturbative grounds,
that the required analyticity hypotheses are indeed satisfied).
By virtue of the {\it optical theorem} and of Eqs. \eqref{scatt-hadron} and
\eqref{scatt-loop}, the total cross section is then given by the expression
\bea
\label{cross-section}
\lefteqn{
\sigma_{\rm tot}^{(hh)} (s) \mathop{\sim}_{s \to \infty}
{\dfrac{1}{s}} {\rm Im} {\cal M}_{(hh)} (s, t=0) }
\nonumber \\
& & = -2\displaystyle\int d^2\vec{R}_{1\perp} \int_0^1 df_1~
|\psi_1(\vec{R}_{1\perp},f_1)|^2
\displaystyle\int d^2\vec{R}_{2\perp} \int_0^1 df_2~
|\psi_2(\vec{R}_{2\perp},f_2)|^2
\nonumber \\
& & \times \int d^2z_{\perp} {\,\rm Re\,}{\cal C}_{E} (\theta \to -i\chi
\simeq -i\log(s/m^2); \vec{z}_{\perp}; \vec{R}_{1\perp},f_1,
\vec{R}_{2\perp},f_2). 
\eea
If one chooses hadron wave functions invariant under rotations and
under the exchange $f_i\to 1-f_i$ (see Refs.~\cite{Dosch,LLCM1} and
also~\cite{pomeron-book}, \S 8.6, and references therein), the
CF ${\cal C}_E$ in Eqs.~(\ref{scatt-hadron}) and
(\ref{cross-section}) can be substituted (without changing the result)
with the following {\it averaged} CF
[$\vec{R}_{i\perp} = |\vec{R}_{i\perp}|(\cos\phi_i,\sin\phi_i)$]:
\bea
\lefteqn{
{\cal C}_E^{ave}(\theta;|\vec{z}_{\perp}|;|\vec{R}_{1\perp}|,f_1,
|\vec{R}_{2\perp}|,f_2) \equiv
\int\frac{d\phi_1}{2\pi} \int\frac{d\phi_2}{2\pi} }
\nonumber 
\label{eq:ave} \\
& & \times\frac{1}{4}\left\{
{\cal C}_E(\theta;\vec{z}_{\perp};\vec{R}_{1\perp},f_1,\vec{R}_{2\perp},f_2)+ 
{\cal C}_E(\theta;\vec{z}_{\perp};\vec{R}_{1\perp},1-f_1,
\vec{R}_{2\perp},f_2)\right. \nonumber \\ 
& & + \left. {\cal C}_E(\theta;\vec{z}_{\perp};\vec{R}_{1\perp},f_1,
\vec{R}_{2\perp},1-f_2) + {\cal C}_E(\theta;\vec{z}_{\perp};
\vec{R}_{1\perp},1-f_1,\vec{R}_{2\perp},1-f_2)\right\}.
\eea
We note here that, as a consequence of the (Euclidean)
{\it crossing--symmetry relations} \cite{crossing},
$\mathcal{C}_E(\pi-\theta;\vec{z}_{\perp};1,2)
=\mathcal{C}_E(\theta;\vec{z}_{\perp};1,\overline{2}) 
=\mathcal{C}_E(\theta;\vec{z}_{\perp};\overline{1},2)$,
where the arguments ``$\overline{i}$'' stand for ``$-\vec{R}_{i\perp}, 1-f_i$''
($i=1,2$), the function ${\cal C}_E^{ave}$ is automatically
{\it crossing--symmetric}, i.e.,
${\cal C}_E^{ave}(\pi-\theta;\ldots)={\cal C}_E^{ave}(\theta;\ldots)$
for fixed values of the other variables.
(The exchange ``$1,2$'' $\to$ ``$1,\overline{2}$'', or ``$1,2$'' $\to$
``$\overline{1},2$'', as well as $\theta \to \pi-\theta$, corresponds to the
exchange from a loop--loop correlator to a loop--{\it antiloop} correlator,
where an {\it antiloop} is obtained from a given loop by exchanging the quark
and the antiquark trajectories.)

\newsection{Wilson--loop correlation functions on the lattice and
comparison with known analytical results}

The gauge--invariant Wilson--loop CF ${\cal C}_E$ is a natural
candidate for a lattice computation.
In Refs. \cite{GM2008,GM2010} a Monte Carlo calculation of ${\cal C}_E$ for
several values of the relative angle and different configurations in the
transverse plane has been performed, using 30000 {\it quenched} configurations
generated with the $SU(3)$ Wilson action at $\beta \equiv 6/g^2 = 6.0$,
corresponding to a lattice spacing $a\simeq 0.1\,{\rm fm}$, on a $16^4$
hypercubic lattice with periodic boundary conditions.
The Wilson--loop CFs have been constructed using loops with transverse sizes
$|\vec{r}_{1\perp}|=|\vec{r}_{2\perp}|=1$ in lattice units
($\vec{R}_{i\perp} = a \vec{r}_{i\perp}$, $\vec{z}_\perp = a \vec{d}_\perp$)
and seven different values of the relative angle $\theta$, i.e.,
$\cot \theta = 0,\pm 1,\pm 2$.
Without loss of generality (see the Appendix of Ref. \cite{GM2010}), the
longitudinal--momentum fractions have been taken to be $f_1=f_2=\frac{1}{2}$:
the loop configurations in the transverse plane that have been studied are
$\vec{d}_{\perp} \parallel \vec{r}_{1\perp} \parallel
\vec{r}_{2\perp}$ (``{\it zzz}'') and $\vec{d}_{\perp} \perp
\vec{r}_{1\perp} \parallel \vec{r}_{2\perp}$ (``{\it zyy}'').
Also the orientation--averaged quantity (``{\it ave}'') defined in
Eq. \eqref{eq:ave} has been measured.
Finally, the CFs have been calculated for the values
$d \equiv |\vec{d}_\perp| = 0,1,2$ of the transverse distance between the
centers of the loops: as expected (see the discussion in Section 4.1 below),
the CFs vanish rapidly as $d$ increases, thus making a ``brute--force''
Monte Carlo calculation very difficult at larger distances. 

As already pointed out in the Introduction, numerical simulations of
LGT can provide the Euclidean CF only for a finite set of
$\theta$--values, and so its analytic properties cannot be directly
attained; nevertheless, they are first--principles calculations that
give us (within the errors) the true QCD expectation for this
quantity. Approximate analytical calculations of this same CF have then
to be compared with the lattice data, in order to test the goodness of
the approximations involved. This can be done either by direct
comparison, when a numerical prediction is available, or by fitting
the lattice data with the functional form provided by a given
model. The Euclidean CFs we are interested in have been evaluated in
the {\it Stochastic Vacuum Model} (SVM)~\cite{LLCM2},
in {\it perturbation theory} (PT) \cite{BB,Meggiolaro05,LLCM2},
in the {\it Instanton Liquid Model} (ILM)~\cite{ILM,GM2010},
and, using the AdS/CFT correspondence, for the ${\cal N}=4$ SYM theory
at large $N_c$, large 't Hooft coupling and large distances between the
loops~\cite{JP}, obtaining, respectively:
\bea
\label{eq:SVM}
{\cal C}^{\,\rm (SVM)}_E(\theta) &=& \frac{2}{3}
\exp\left[-\frac{1}{3} K_{\rm SVM}\cot\theta\right]
+ \frac{1}{3} \exp\left[\frac{2}{3} K_{\rm SVM}\cot\theta\right] - 1,\\
\label{eq:pert}
{\cal C}_E^{(\rm PT)}(\theta) &=& K_{\rm PT} \cot^2\theta,\\
\label{eq:ILM}
{\cal C}^{\,\rm (ILM)}_E(\theta) &=& \frac{K_{\rm ILM}}{\sin\theta},\\
\label{eq:AdS}
{\cal C}^{\,\rm (AdS/CFT)}_E(\theta) &=&
\exp\left[\dfrac{K_1}{\sin\theta} + K_2\cot\theta +
K_3\cos\theta\cot\theta\right]-1 ,
\eea 
where the coefficients $K_i = K_i(\trpl{z};1,2)$ are functions of
$\trpl{z}$ and of the dipole variables $\vec{R}_{i\perp}, f_i$.
The comparison of the lattice data with these analytical calculations is not,
generally speaking, fully satisfactory.
The values of the chi--squared per degree of freedom ($\chi^2_{\rm d.o.f.}$)
of the best fits, performed in Ref. \cite{GM2008} using the above--reported
functions \eqref{eq:SVM}--\eqref{eq:AdS}, are listed in Table \ref{tab:chi2}
(together with the values obtained from best fits with the parameterizations
\virg{Corr 1}, \virg{Corr 2} and \virg{Corr 3}, that we shall introduce and
discuss in the next section).
As one can see in Table \ref{tab:chi2},
largely improved best fits have been obtained by combining the ILM and
perturbative expressions into the following expression:
\be
\label{eq:ILMp}
{\cal C}^{\,\rm (ILMp)}_E(\theta) = \frac{K_{\rm ILMp1}}{\sin\theta}
+ K_{\rm ILMp2}\cot^2\theta .
\ee
As we have said in the Introduction, the main motivation in studying
soft high--energy scattering is that it can lead to a resolution of
the total cross section puzzle. From this point of view, the analytical
models considered in this section are absolutely unsatisfactory, since they
do not lead to rising, or, better, to {\it Froissart--like} total cross sections
of the form $B \log^2 s$ at high energy, as experimental data seem
to suggest. In fact, the SVM, PT, ILM and ILMp parameterizations
\eqref{eq:SVM}--\eqref{eq:ILM} and \eqref{eq:ILMp} lead to asymptotically
constant total cross sections\footnote{Actually the ILM parameterization
\eqref{eq:ILM} leads to null total cross sections!},
as it can be seen by using Eq.~\eqref{cross-section}.
Concerning the AdS/CFT expression \eqref{eq:AdS}, obtained in ${\cal N}=4$ SYM,
it has been shown in~\cite{GP2010} that, by combining the knowledge of
the various coefficient functions $K_i$ in \eqref{eq:AdS} at large
$|\trpl{z}|$~\cite{JP} with the unitarity constraint in the small--$|\trpl{z}|$
region, a non--trivial high--energy behavior for the {\it dd} total cross
section can emerge (including a {\it Pomeron}--like behavior
$\sigma \sim s^{1/3}$: note that, since a Conformal Field Theory has no
\emph{mass gap}, there is no need for the \emph{Froissart bound} to hold
also in this case).
\begin{table}[t]
\centering
%{\footnotesize 
 \begin{tabular}[h]{|l|cc|ccc|ccc|}
\hline
$\chi^2_{\rm d.o.f.}$ & \multicolumn{2}{c|}{$d=0$} &
\multicolumn{3}{c|}{$d=1$} & \multicolumn{3}{c|}{$d=2$}\\
& {\it zzz}/{\it zyy} & {\it ave} & {\it zzz} & {\it zyy} &
{\it ave} & {\it zzz} & {\it zyy} & {\it ave} \\
\hline
SVM & 51 & - & 16 & 12 & - & 1.5 & 2.2 & -\\
PT & 53 & 34 & 16 & 13 & 13 & 1.5 & 2.2 & 4.5\\
ILM & 114 & 94 & 14 & 15 & 45 & 0.45 & 0.35 & 1.45\\
ILMp & 20 & 9.4 & 0.54 & 0.92 & 1.8 & 0.13 & 0.12 & 0.19\\
AdS/CFT & 40 & - & 1 & 0.63 & - & 0.14 & 0.065 & -\\
\hline
Corr 1 & 12.9 & 2.81 & 0.34 & 0.66 & 1.25 & 0.16 & 0.07 & 0.05 \\
Corr 2 & 7.88 & 0.55 & 0.27 & 0.55 & 0.31 & 0.15 & 0.07 & 0.05 \\
Corr 3 & 3.89 & 0.17 & 0.16 & 0.77 & 0.11 & 0.12 & 0.10 & 0.10 \\
\hline
\end{tabular}
%}
\caption{Chi--squared per degree of freedom for a best fit with the indicated
function: SVM [Eq. \eqref{eq:SVM}], PT [Eq. \eqref{eq:pert}],
ILM [Eq. \eqref{eq:ILM}], ILMp [Eq. \eqref{eq:ILMp}],
Corr 1 [Eq. \eqref{Corr1E}], Corr 2 [Eq. \eqref{Corr2E}],
Corr 3 [Eq. \eqref{Corr3E}].}
\label{tab:chi2}
\end{table}

\newsection{How a Froissart--like total cross section can be obtained:
a new analysis of Wilson--loop correlators from lattice QCD}

An ambitious
question that one can ask at this point is if the lattice data are
compatible with rising total cross sections. An answer can in
principle be obtained by performing best fits to the lattice data
with more general functions, leading to a non--trivial dependence on
energy. This approach requires special care, because of the analytic
continuation necessary to obtain the physical amplitude from the
Euclidean CF: one has therefore to restrict the set of admissible
fitting functions by imposing physical constraints, first of all
{\it unitarity}.

Introducing the \virg{\emph{hadron--hadron correlator}} $\mathcal{C}^{(hh)}_M$
as the average of the \virg{\emph{dipole--dipole correlator}} $\mathcal{C}_M$
over the dipole variables, weighted with the proper squared hadronic wave
functions, i.e.,
\bea
\mathcal{C}^{(hh)}_M(\chi;|\trpl{z}|) &\equiv& \wfint \nonumber \\
&\times& \mathcal{C}_M(\chi;\trpl{z};\mvar{1},\mvar{2}) ,
\label{corrhh}
\eea
one immediately recognizes from Eqs. \eqref{scatt-hadron} and
\eqref{scatt-loop} (see Refs. \cite{BN} and \cite{LLCM1}) that
$\mathcal{C}^{(hh)}_M(\chi;|\trpl{z}|)$ is nothing but
the scattering amplitude $A(s,|\trpl{z}|)$ in impact--parameter space
(i.e., the partial--wave scattering amplitude $S_l-1=\eta_l e^{2i\delta_l}-1$),
which must satisfy the following well--known unitarity condition (see, for
example, Refs. \cite{unitarity}): $|A(s,|\trpl{z}|)+1| \leq 1$.
Therefore, this unitarity condition immediately translates to:
\begin{equation}\label{unitcorrhh}
\big|\mathcal{C}^{(hh)}_M(\chi;|\trpl{z}|)+1\big| \leq 1 .
\end{equation}
%(We have used the fact that the hadron--hadron correlator
%$\mathcal{C}^{(hh)}_M$ depends only on the modulus of $\trpl{z}$, for the
%invariance under rotations in the transverse plane.)
Since the hadronic wave functions are normalized to 1, the unitarity condition
(\ref{unitcorrhh}) is obviously satisfied if the following \emph{sufficient}
(and therefore \virg{\emph{stronger}}) condition for the loop--loop correlator
$\mathcal{C}_M(\chi;\trpl{z};1,2) $ holds:
\begin{equation}\label{unitcorrdd}
\big|\mathcal{C}_M(\chi;\trpl{z};\mvar{1},\mvar{2})+1\big| \leq 1 \qquad
\forall \,\,\trpl{z},~\trpl{R_1},~\trpl{R_2},~f_1,~f_2 ,
\end{equation}
i.e., if the dipole--dipole correlator stays inside the \emph{Argand circle}
for all values of $\trpl{z} $, $\trpl{R_1} $, $\trpl{R_2} $, $f_1$, $f_2$.
It is also possible to find another (\virg{\emph{weaker}}) \emph{sufficient}
unitarity condition,
in terms of the averaged correlator $\mathcal{C}_M^{ave}$,
which is the Minkowskian version of $\mathcal{C}_E^{ave}$, defined in
Eq. (\ref{eq:ave}) in the Section 2:
in fact, as we have said there, we can substitute the correlator
$\mathcal{C}_M$ with the \emph{averaged} correlator $\mathcal{C}_M^{ave}$
(without changing the result), whenever it is averaged over the dipole
variables $\trpl{R_i} $ and $f_i$ (with $i=1,2$) with the proper (squared)
hadronic wave functions $|\psi_1|^2$ and $|\psi_2|^2$, as, for example,
in Eq. (\ref{corrhh}). Therefore, the unitarity condition (\ref{unitcorrhh})
is also satisfied if the following \emph{sufficient} condition for the averaged
loop--loop correlator (\emph{stronger} than (\ref{unitcorrhh}), but
\emph{weaker} than (\ref{unitcorrdd})) holds:
\begin{equation}\label{unitcorrave}
\big|\mathcal{C}^{ave}_M(\chi;|\trpl{z}|;|\trpl{R_1}|,f_1,|\trpl{R_2}|,f_2)+1
\big| \leq 1 \qquad
\forall \,\,|\trpl{z}|,~|\trpl{R_1}|,~|\trpl{R_2}|,~f_1,~f_2.
\end{equation}

\newsubsection{General considerations on the form of the correlator}

In this section, we are going to introduce, and partially justify, new
parameterizations of the CF that, in order:
i) fit well the data;
ii) satisfy the unitarity condition after analytic continuation; and
iii) lead to total cross sections rising as $B \log^2 s$ in the high--energy
limit (as experimental data seem to suggest).\\
We will show that the above conditions lead to rather strong constraints about
the possible shape of the parameterization.
First of all, we observe that the conditions ii) and iii) cannot be
simultaneously satisfied by a correlator with the following
factorized form: $-\re\, \mathcal{C}^{(hh)}_M(\chi;|\trpl{z}|)= \varsigma(\chi)
\upsilon(|\trpl{z}|)$,
with $\varsigma(\chi)$ rising with $\chi$ for $\chi\to\infty$, since the
unitarity condition (\ref{unitcorrhh}) implies that
$0\leq-\re \mathcal{C}^{(hh)}_M(\chi;|\trpl{z}|)\leq2 $ (actually, with this
factorized form, one can only have $\sigma^{(hh)}_{\text{tot}}(\chi)\to
\text{constant}$ for $\chi\to\infty$).
This means that rising total cross sections can be obtained, without violating
unitarity, only if the correlator is {\it not} factorizable. Let us now give
a few general considerations about the form of the Euclidean loop--loop
correlator. As a starting point, we shall \emph{assume} that the Euclidean
correlator can be written as:
\begin{equation}\label{shapeE}
\mathcal{C}_E(\theta;\trpl{z};1,2)=\exp\big[K_E(\theta;\trpl{z};1,2)\big]-1,
\end{equation}
where $K_E(\theta;\trpl{z};1,2)$ is a \emph{real} function (since the correlator
$\mathcal{C}_E$ itself is known to be a \emph{real} function \cite{GM2008}).
This assumption, i.e., essentially the fact that: $\mathcal{C}_E+1\geq0$, is
indeed rather well justified for many reasons: first,
in the large--$N_c$ expansion, the correlator $\mathcal{C}_E$ is expected
to be of order $\mathcal{O}(1/N_c^2)$ (see Eq. (3.4) in Ref. \cite{GM2008}),
so that $\mathcal{C}_E+1\geq0$ is certainly satisfied for large $N_c$;
moreover, all the known analytical models (SVM, ILM, AdS/CFT correspondence,
perturbation theory,\dots) actually satisfy it; and last (but not least!),
the lattice data obtained in Refs. \cite{GM2008,GM2010} confirm
it.\footnote{Actually, lattice data
\cite{GM2008} seem to support an even {\it stronger} condition, i.e., the
positivity of the correlator itself, $\mathcal{C}_E\geq0$, which, in terms of
the parameterization (\ref{shapeE}) would mean $K_E\geq0$. We are not aware of
any rigorous proof of this condition, apart from a rough argument in terms of
\virg{minimal surfaces}. (Let us observe also that this {\it stronger}
condition is {\it not} expected to hold in the Abelian case: in fact, the
parameterization for the Euclidean correlator ${\cal C}_E$ in {\it quenched}
QED, found in Ref. \cite{Meggiolaro05}, while still being of the form
\eqref{shapeE}, has not the property $K_E \geq 0$.)}\\
At this point, the Minkowskian correlator can be obtained after analytic
continuation:
\begin{equation}\label{shapeM}
\mathcal{C}_M(\chi;\trpl{z};1,2)=\exp\big[K_M(\chi;\trpl{z};1,2)\big]-1,
\end{equation}
with $K_M(\chi;\trpl{z};1,2) = K_E(\theta \to -i\chi;\trpl{z};1,2)$.
In the large--$\chi$ limit, the Minkowskian correlator $\mathcal{C}_M$ is expected to obey the unitarity condition (\ref{unitcorrdd}), which using the parameterization (\ref{shapeM}), reduces to the following very simple relation:
\begin{equation}\label{unitcorrKM}
\re\, K_M(\chi;\trpl{z};1,2)\leq0\quad \forall
\trpl{z}, \trpl{R_i}, f_i\quad (i=1,2) .
\end{equation}
%(this is nothing but the condition on the \virg{\emph{inelasticity parameter}}
%in the partial--wave representation).
So, the parameterizations that we are going to consider have the general form
\begin{equation}\label{corrEushape}
\mathcal{C}_E(\theta;\trpl{z};1,2)=\exp\Big[\sum_{i} K_i(\trpl{z};1,2)
F_{Ei}(\theta)\Big]-1,
\end{equation}
where the sum is over different terms with various functions $F_{Ei}(\theta)$
and various \virg{parameters} $K_i(\trpl{z};1,2)$.
We now make an important consideration about the dependence of the parameters
$K_i$ on the impact parameter $|\trpl{z}|$.
For a {\it confining} theory like QCD the loop--loop correlator
$\mathcal{C}_E$ is expected to decay exponentially at large $|\trpl{z}|$ as
\begin{equation}\label{corrlargez}
\mathcal{C}_E\sim\alpha\,e^{-\mu |\trpl{z}|},
\end{equation}
where $\mu$ is some mass--scale proportional to the mass of the lightest
glueball ($M_G\simeq1.5$ GeV) or maybe (as suggested, for example, by the
SVM model: see below) to the inverse $1/\lambda_{vac}$ of the so--called
\emph{vacuum correlation length} $\lambda_{vac}$, which has been measured
with Monte Carlo simulations on the lattice in Refs. \cite{lambda-vacuum}
(see also Ref. \cite{DDSS} for a review),
both in the \emph{quenched} ($\lambda_{vac}\simeq0.22$ fm)
and \emph{full} QCD ($\lambda_{vac}\simeq0.30$ fm).
(For example, in the SVM model, see Eq. (\ref{eq:SVM}), one finds
\cite{LLCM2} that, for $|\trpl{z}|\to\infty$,
$K_{\text{SVM}}\sim\,e^{-|\trpl{z}|/\lambda_{vac}}$, so that
$\mathcal{C}_E^{\text{SVM}} \sim
\frac{1}{9}K_{\text{SVM}}^2 \cot^2\theta\sim\alpha\,e^{-\mu |\trpl{z}|}$,
with $\mu=\frac{2}{\lambda_{vac}}$.)
Therefore, we should require the same large--$|\trpl{z}|$ behavior
\eqref{corrlargez} for the parameters $K_i$ in Eq. \eqref{corrEushape},
i.e., $K_i \sim e^{-\mu |\trpl{z}|}$.
Instead, for a non--confining theory, e.g., for a \emph{conformal} field theory,
also different behaviors of the parameters $K_i$ for large $|\trpl{z}|$ are
possible, typically like powers of $1/|\trpl{z}|$. This is what happens for the
parameters $K_{1,2,3}$ for the parameterization (\ref{eq:AdS})
obtained from the AdS/CFT correspondence \cite{JP,GP2010}.
However, it can be shown that, in the general case, an asymptotic
large--$|\trpl{z}|$ behavior of the parameters $K_i$ like powers of
$1/|\trpl{z}|$ leads to \emph{non--universal} high--energy total cross sections,
and can reproduce a Froissart--like behavior,
$\sigma^{(hh)}_{\text{tot}}\sim B \log^2 s$, only with very \virg{\emph{ad hoc}}
dependencies of the parameters $K_i$ on powers of $1/|\trpl{z}|$ and of the
functions $F_{Ei}(\theta)$ on powers of $\theta$.
%This and other more technical details and generalizations will be
%discussed in a separate publication \cite{GMM}.

\newsubsection{How a Froissart--like total cross section can be obtained}

Let us now assume that the leading term for $\chi\to+\infty$ (i.e., for
$s\to\infty$) of the Minkowskian dipole--dipole CF is of the form
\begin{equation}\label{corrasym}
\mathcal{C}_M(\chi;\trpl{z};1,2)\mathop{\sim}_{\chi\to+\infty}
\exp\big(i\, \beta\,f (\chi)\,e^{-\mu |\trpl{z}|}\big)-1,
\end{equation}
where $\beta=\beta(1,2)$ is a function of the dipole variables and $f(\chi)$
is a \emph{positive} and \emph{real} function rising with $\chi$, i.e,
$f(\chi)\to+\infty$ for $\chi\to +\infty$: for example we can have
$f(\chi)=e^{n\chi},\,(\cosh\chi)^n,\, \chi^p e^{n\chi},\dots$. It is then
clear that, in order to satisfy the unitarity condition (\ref{unitcorrdd}),
the imaginary part of $\beta$ has to be positive, i.e,
\begin{equation}\label{betaunit}
|\mathcal{C}_M(\chi;\trpl{z};1,2)+1|\leq 1 \quad \Leftrightarrow \quad
\im\,\beta\geq0.
\end{equation}
The precise $\trpl{z}$ dependence of (\ref{corrasym}) is, of course,
expected to be valid only for large enough $|\trpl{z}|$. For simplicity,
we shall first assume that it is valid $\forall\,|\trpl{z}|\geq0$.\\
Following Eq. (\ref{cross-section}), and performing the change of variable
$y=\mu |\trpl{z}|$, we have that:
\begin{equation}\label{sigmaequation}
\sigma^{(hh)}_{\text{tot}} \sim
\frac{4\pi}{\mu^2}\re\wfint I(\chi,\beta),
\end{equation}
where the quantity $I(\chi,\beta)$ is defined as
\begin{equation}\label{integral}
I(\chi,\beta) \equiv \int_0^\infty dy \, y\big[1-\exp\big(i\,\beta f(\chi)
e^{-y}\big)\big] .
\end{equation}
We can now expand the exponential in series, and exchange the order of
integration and summation, obtaining:
\begin{equation}\label{series1}
I(\chi,\beta)=-\sum_{n=1}^\infty \frac{(i\beta f(\chi))^n}{n!}
\int_0^\infty dy \, y\, e^{-ny} = -\sum_{n=1}^\infty
\frac{(i\beta f(\chi))^n}{n!n^2}.
\end{equation}
%Note that, if we make the substitution $N=n-1$, we obtain
%\begin{eqnarray}\label{series2}
%\nonumber I(\chi,\beta)&=&-\sum_{N=0}^\infty \frac{(i\beta f(\chi))^{N+1}}{(N+1)^2(N+1)!}=-(i\beta f(\chi))\sum_{N=0}^\infty \frac{(i\beta f(x))^N}{(N+1)^3N!},\\
%&=& -i\beta f(x) {}_3F_3(1,1,1;2,2,2;i\beta f(\chi)) ,
%\end{eqnarray}
%where (see, e.g., Ref. \cite{GR})
%\begin{equation}
%_p F_q(a_1,\dots,a_p;b_1,\dots,b_q;z)=\sum_{n=0}^\infty \frac{(a_1)_n\ldots (a_p)_n}{(b_1)_n\ldots (b_q)_n} \frac{z^n}{n!}
%\end{equation}
%is the Generalized Hypergeometric Function of kind $(p,q)$, and $(a)_n= \frac{(a+n-1)!}{a!}$ is the Pochhammer symbol. Note that the power series converges uniformly (with infinite radius of convergence) since, defining $\alpha_N \equiv \frac{1}{(N+1)^3N!}$, we have $\mathop{\lim}\limits_{N\to\infty}\frac{\alpha_{N+1}}{\alpha_N}=0$.\\
The expression \eqref{series1}
%--\eqref{series2}
is valid for an arbitrary function $f(\chi)$.
However, as we have said, we are interested only in its asymptotic form
for large $\chi$, in the case of $f(\chi)$ rising with $\chi$,
i.e., $f(\chi)\to+\infty$ for $\chi\to +\infty$.
Thus $\exists\, \chi_0\in \mathbb{R}^+$ s.t. $f(\chi)>0$, $\forall\,
\chi\geq \chi_0$. So, for $\chi\geq\chi_0$, we can define the variable
$\eta\equiv\log f(\chi)$, and re--write the quantity $I(\chi,\beta)$, Eqs.
\eqref{integral}--\eqref{series1}, as a function of $\eta$ and $\beta$:
\begin{equation}\label{inteta}
I(\chi,\beta) = J(\eta,\beta) \equiv
%\int_0^\infty dy \, y \big[1-\exp\big(i\,\beta e^{\eta-y}\big)\big] =
-\sum_{n=1}^\infty \frac{(i\beta e^\eta)^n}{n!n^2}.
\end{equation}
%Note that $J$ has the following symmetry property with respect to its
%variables:
%\begin{equation}\label{intsym}
%J(\eta+A,\beta) = J(\eta,\beta\,e^A), \quad \forall A\in\mathbb{R}.
%\end{equation}
Now, deriving (\ref{inteta}) with respect to the variable $\eta$, one finds
\begin{equation}
\frac{\partial J}{\partial \eta}=-\sum_{n=1}^{\infty} \frac{(i\beta\,e^\eta)^n}{n! n}.
\end{equation}
The sum of the above series is known (see, e.g, Ref. \cite{GR}) and given by
\begin{equation}\label{intprimo}
J'(\eta,\beta)=E_1(-i\beta\,e^\eta) + \log(-i\beta\,e^\eta) + \gamma,
\quad -\pi<\arg(-i\beta e^\eta)<\pi,
\end{equation}
where $\gamma$ is the Euler--Mascheroni constant ($\gamma\simeq0.57721\dots$)
and $E_1(z)$ is the Schloemilch's exponential integral
(see, e.g., Ref. \cite{GR}).
%\begin{equation}
%E_1(z)=\int_1^\infty \frac{e^{-tz}}{t}dt, \quad -\pi<\arg(z)<\pi.
%\end{equation}
Since $E_1(z) \sim e^{-z}/z$ at large $|z|$, for $\re \, z \ge 0$,
%\begin{equation}
%E_1(z)\mathop{\sim}_{|z|\gg1} \frac{e^{-z}}{z}
%\bigg[1-\frac{1}{z}+\mathcal{O}\bigg(\frac{1}{z^2}\bigg)\biggr],
%\quad \re\, z\geq0.
%\end{equation}
and, moreover,
$\re (-i\beta\,e^\eta) \geq 0 \Leftrightarrow \im \beta \geq 0$
is nothing but the unitarity condition (\ref{betaunit}),
%we can conclude that:
%$E_1(-i\beta\,e^\eta)\sim\mathcal{O}(e^{-\eta})$ for $\eta\to+\infty$.
the asymptotic form of (\ref{intprimo}) is readily obtained:
\begin{equation}\label{intprimoasy}
J'(\eta,\beta) \mathop{\sim}_{\eta\to+\infty}
\eta + \log(-i\beta) + \gamma + \mathcal{O}(e^{-\eta}).
\end{equation}
Now we can re--integrate (\ref{intprimoasy}) in $\eta$, finding
\begin{equation}\label{intfinaleeta}
J(\eta,\beta)\mathop{\sim}_{\eta\to+\infty}\frac{\eta^2}{2} +
\eta \big(\log(-i\beta) + \gamma\big) + \text{constant} +
\mathcal{O}(e^{-\eta}),
\end{equation}
and, making the substitution $\eta = \log f(\chi)$,
\begin{equation}\label{intfinalex}
I(\chi,\beta)\mathop{\sim}_{\chi\to+\infty}\frac{\log^2 f(\chi)}{2}+\log f(\chi)
\big(\log(-i\beta) + \gamma\big) + \text{constant} +
\mathcal{O}\bigg(\frac{1}{f(\chi)}\bigg).
\end{equation}
Let us observe the following important fact:
the leading order in $\chi$ (and $\eta$) does \emph{not} depend on $\beta$.
So, coming back to Eq. (\ref{sigmaequation}), the asymptotic behavior of the
total cross section turns out to be:
\begin{multline}
\label{intwithmandn}
~~~~~~~~~\sigma^{(hh)}_{\text{tot}} \sim
\frac{4\pi}{\mu^2}\wfint\\
\times \bigg[\frac{\log^2 f(\chi)}{2}+\log f(\chi)
\big(\log|\beta| + \gamma\big)+\dots\bigg],~~~~~~~~~~~~~~~~~~~~~
\end{multline}
where the whole dependence on the dipole variables
($\trpl{R_i} $, $f_i $, $i=1,2$) is in the coefficient $\beta$.
If one assumes a leading term of the type $f(\chi)= e^{n\chi}$ or, even more
generally, $f(\chi)= \chi^p e^{n\chi} $ in the correlator, the resulting
asymptotic behavior for the total cross section is (recalling that
$\chi\simeq\log(s/m^2)$):
\begin{equation}\label{sigmatotlead}
\sigma^{(hh)}_{\text{tot}} \sim B \log^2 s,
\qquad \text{with:}\qquad B = \frac{2\pi n^2}{\mu^2}.
\end{equation}
We want to emphasize the fact that the above result is \emph{universal},
depending only on the mass scale $\mu$, which, as we have said, sets the
large--$|\trpl{z}|$ dependence of the correlator. In fact, the integration over
the dipole variables does not affect the coefficient of the leading term, since
the hadronic wave functions are normalized to 1.
Let us also observe, that the \emph{universal} coefficient $B$ is not affected
by the masses of the scattering particles. In fact, in the case of scattering
of two different mesons with masses $m_1$ and $m_2$, the rapidity is
$\chi\sim\log(\frac{s}{m_1 m_2})$. Therefore, considering the case of two
different mesons (or, in general, a change of the energy scale implicitly
contained in Eq. \eqref{sigmatotlead}) does not affect the \emph{universal}
coefficient $B$ of the leading term (but will in general affect sub--leading
$\log$ and {\it constant} terms), since
$\chi\sim\log(\frac{s}{m_1 m_2})=\log(\frac{s}{s_0})+\log(\frac{s_0}{m_1 m_2})$,
with $\sqrt{s_0}$ being an arbitrary energy scale.
This is the main theoretical achievement in this work.

This same relation can be derived also with less stringent conditions on the
$\trpl{z}$ dependence, assuming (as it must be!) the exponential--type
dependence in Eqs. (\ref{corrlargez}) and (\ref{corrasym})
only for $|\trpl{z}| > z_0$, with $z_0$ much larger than
$1/\mu$ and the dipole sizes.
Starting from the expression (\ref{cross-section}) of the total cross section,
re--written as\\
$\sigma^{(hh)}_{\text{tot}}=-4\pi\re\int_0^\infty |\trpl{z}|\,d|\trpl{z}| \,
\mathcal{C}^{(hh)}_M(\chi;|\trpl{z}|)$,
where $\mathcal{C}^{(hh)}_M(\chi;|\trpl{z}|)$
has been defined in Eq. (\ref{corrhh}),
one can split the integration in the variable $|\trpl{z}|$ in two parts,
a \virg{\emph{tail}} contribution ($\int_{z_0}^\infty d|\trpl{z}|\dots$),
which can be evaluated using the approximate expression (\ref{corrasym}) for
the loop--loop correlator, and a \virg{\emph{core}} contribution
($\int_0^{z_0}d|\trpl{z}|\dots$), which, instead, can be bounded using the
unitarity condition \eqref{unitcorrhh}
[or \eqref{unitcorrdd}--\eqref{unitcorrave}], in the form:
$-2\leq\re\,\mathcal{C}_M \leq 0$.
%This and other more technical details and generalizations will be
%discussed in a separate publication \cite{GMM}.

In the next subsection we are going to show our new analysis of the lattice
data.

\newsubsection{New parameterizations for the correlator}

In what follows we show some parameterizations that we have found that satisfy
the criteria i)--iii) listed above, together with the corresponding estimate
of the asymptotic term of the high--energy total cross sections.
Despite the appearances, it has not been simple to find such parameterizations:
what follows is a selection between more than 70 different
parameterizations that we have tried. For each proposed parameterization of the
correlator, we are going to show the $\chi_{d.o.f.}^2$ of the corresponding
best fit to the lattice data for each given transverse distance and
each given configuration: the results are summarized in Table \ref{tab:chi2}.
As we have already pointed out in Section 2, the averaged correlator
$\mathcal{C}^{ave}$ is somehow \virg{closer} to the hadron--hadron scattering
matrix ${\cal M}_{(hh)}$ than the correlator at fixed transverse configuration
(like \virg{$zzz$} or \virg{$zyy$}), since it is actually the result of an
integration of the dipole--dipole correlator over the orientations of the
dipoles. The analysis performed above in Sections 4.1 and 4.2 can be repeated
for the averaged correlator $\mathcal{C}^{ave}$ without altering any
conclusions. For this reason we are going to focus our analysis on the averaged
correlator $\mathcal{C}^{ave}$ only; however, the $\chi_{d.o.f.}^2$ of the
best fits to the \virg{$zzz$} and the \virg{$zyy$} data are also shown for
comparison in Table \ref{tab:chi2}. Since, as we have already said in
Section 2, the averaged correlator $\mathcal{C}^{ave}$ is automatically
\emph{crossing--symmetric}, so are the parameterizations that we
propose.\footnote{However, it was observed in Ref. \cite{GM2008} that a small
(but nonzero!) \emph{Odderon} ($C$--odd) contribution in {\it dd} scattering,
which is related through the {\it crossing--symmetry relations}
\cite{crossing} to the antisymmetric part of ${\cal C}_E(\theta)$
with respect to $\theta=\frac{\pi}{2}$, is present in the lattice data
corresponding to the ``$zzz$'' and ``$zyy$'' transverse configurations.
As noticed there, a \emph{crossing--antisymmetric} term in the exponent $K_E$
of the Euclidean CF \eqref{shapeE} proportional to $\cot\theta$ (as,
for example, the one appearing in the SVM parameterization (\ref{eq:SVM}) and
also the one appearing in the AdS/CFT parameterization (\ref{eq:AdS})) is in
general suitable for taking into account this {\it Odderon} contribution and
fits quite well the lattice data for the antisymmetric part of the correlator.
Let us note also that its analytic continuation, $\cot\theta \to i\coth\chi$,
is limited for $\chi\to\infty$, and so it is consistent with the
\emph{Pomeranchuk theorem}, at least for rising total cross sections.}\\
With regard to the explicit angular dependence of the possible terms
$F_{Ei}(\theta)$, let us make some preliminary considerations. It is known
\cite{GM2008,GM2010} that lattice data for the correlator blow up at
$\theta=0^\circ,180^\circ$ (as expected from the relation between the correlator
and the dipole--dipole static potential \cite{Pot}), and that they are clearly
different from zero for $\theta=90^\circ$. A simple term like $1/\sin\theta$
(which always comes out as a Jacobian in the integration over the longitudinal
coordinates in the analytical models considered in the previous section) can
account for such a behavior and, as noticed in Refs. \cite{GM2008,GM2010},
it fits quite well the data around $\theta=90^\circ$. Therefore, we shall
always include a $1/\sin\theta$ term in our parameterizations for the exponent
in Eq. (\ref{corrEushape}).\\
Concerning the analysis of the impact--parameter dependence, we want to stress
the fact that it must be taken only as an estimate, since only a few (small)
values of the impact parameter are available from the lattice data
($|\trpl{z}|=a d$, with $d=0,1,2$).

\newsubsubsection{Correlator 1}

Following a first possible strategy, we have tried to improve best fits
achieved with the ILMp expression \eqref{eq:ILMp}: the idea is to combine
known QCD results and variations thereof. As an example, one could
consider exponentiating the two--gluon exchange and the one--instanton
contribution (i.e., the ILMp expression), and supplementing it with a
term which could yield a rising cross section, e.g., a term proportional to
$\cos\theta\cot\theta$, like the one present in the AdS/CFT parameterization
\eqref{eq:AdS}. We thus find the following parameterization:
\begin{equation}\label{Corr1E}
\mathcal{C}_E(\theta)=\exp\bigg[\frac{K_1}{\sin\theta}+K_2 \cot^2\theta +
K_3 \cos\theta\cot\theta\bigg]-1,
\end{equation}
whose Minkowskian counterpart is:
\begin{equation}\label{Corr1M}
\mathcal{C}_M(\chi)=\exp\bigg[i\,\Big(\frac{K_1}{\sinh\chi}+
K_3 \cosh\chi\coth\chi\Big)-K_2 \coth^2\chi\bigg]-1.
\end{equation}
The unitarity condition \eqref{unitcorrKM} is satisfied if $K_2\geq0$:
from Table \ref{tab:corr123} one sees that the parameter $K_2$ obtained from
a best fit satisfies this condition, within the errors.
\begin{table}
\centering
$\begin{array}{|c|c|c|c|}
\hline
\text{Corr 1} & d=0 & d=1 & d=2
\\\hline
{ K_1} & 5.85(42)\cdot 10^{-3} & 3.07(37)\cdot 10^{-3}
& 8.7(3.1)\cdot 10^{-4} \\
{ K_2} & 9.60(98)\cdot 10^{-2} & 2.44(49)\cdot 10^{-2}
& -5.3(84.5)\cdot 10^{-5} \\
{ K_3} & -7.8(1.3)\cdot 10^{-2} & -1.37(72)\cdot 10^{-2}
& 1.7(1.9)\cdot 10^{-3} \\
\chi_{\text{d.o.f.}}^2 & 2.81 & 1.25 & 0.05 \\ \hline
\text{Corr 2} & d=0 & d=1 & d=2
\\ \hline
{ K_1} & 6.03(42)\cdot 10^{-3} & 3.26(38)\cdot 10^{-3}
& 8.7(3.2)\cdot 10^{-4} \\
{ K_2} & 4.63(46)\cdot 10^{-1} & 1.33(25)\cdot 10^{-1}
& -1.2(54.2)\cdot 10^{-4} \\
{ K_3} & -4.54(50)\cdot 10^{-1} & -1.26(28)\cdot 10^{-1}
& 1.7(6.7)\cdot 10^{-3} \\
\chi_{\text{d.o.f.}}^2 & 0.55 & 0.31 & 0.05 \\ \hline
\text{Corr 3} & d=0 & d=1 & d=2
\\ \hline
{ K_1} & 6.02(36)\cdot 10^{-3} & 3.46(29)\cdot 10^{-3}
& 1.07(20)\cdot 10^{-3} \\
{ K_2} & 1.29(5)\cdot 10^{-1} & 4.47(27)\cdot 10^{-2}
 & 2.11(73)\cdot 10^{-3} \\
\chi_{\text{d.o.f.}}^2 & 0.17 & 0.11 & 0.10 \\ \hline
\end{array}$
\caption{Parameters (with their errors) for the Correlators 1 [Eq.
\eqref{Corr1E}], 2 [Eq. \eqref{Corr2E}], and 3 [Eq. \eqref{Corr3E}], obtained
from best fits to the averaged lattice data, and the corresponding
$\chi^2_{\text{d.o.f.}}$, for the transverse distances $d=0,1,2$.}
\label{tab:corr123}
\end{table}
The best--fit functions are plotted in Fig. \ref{corr1fit}.
Performing a best fit with an exponential function
$\sim e^{-\mu |\trpl{z}|} $ over the three distances, one finds that the
coefficient $K_3$ of the leading term for $\chi\to\infty$ has a mass--scale
$\mu=4.64(2.38)\,\text{GeV}$,
that, following the result (\ref{sigmatotlead}) of the previous section
(with $n=1$, as implied by this parameterization), leads to the following
asymptotic total cross section:
\begin{equation}\label{sigmatotcorr1}
\sigma^{(hh)}_{\text{tot}} \sim B \log^2 s,
\qquad \text{with:}\qquad B = 0.113^{+0.364}_{-0.037} \,\,\, \text{mb}.
\end{equation}
This is compatible, within the large errors, with the experimental result
$B_{exp} \simeq 0.3$ mb reported in the Introduction.

\newsubsubsection{Correlator 2}

Another possible strategy is suggested again by the AdS/CFT expression
\eqref{eq:AdS}: one can try to adapt to the case of QCD the analytical
expressions obtained in related models, such as ${\cal N}=4$ SYM.
Although, of course, Eq.~\eqref{eq:AdS} is not expected to describe QCD,
it is sensible to assume in this case a similar functional form
(basically assuming the existence of the yet unknown gravity dual for
QCD). Assuming moreover that the known power--law behavior of the $K_i$'s
(expected for a conformal theory) goes over into an exponentially
damped one (expected for a confining theory), $K_i \sim e^{-\mu |\trpl{z}|}$,
one obtains a Froissart--like total cross section
$\sigma^{(hh)}_{\text{tot}} \sim B \log^2 s$.
In this spirit, the second parameterization that we propose is:
\begin{equation}\label{Corr2E}
\mathcal{C}_E(\theta)=\exp\bigg[\frac{K_1}{\sin\theta} +
K_2 \left(\frac{\pi}{2}-\theta\right) \cot\theta +
K_3 \cos\theta\cot\theta\bigg]-1.
\end{equation}
This one contains, in addition to the usual AdS/CFT--like terms $1/\sin\theta$,
$\cot\theta$ and $\cos\theta\cot\theta$, also another term proportional to
$\theta\cot\theta$:\footnote{Although such a term could seem \virg{strange}
at first sight, not being a \virg{simple} combination of $\sin\theta$ and
$\cos\theta$, it has been shown in the first Ref. \cite{analytic}, through an
explicit calculation up to the order $\mathcal{O}(g^4)$ in perturbation theory,
that a similar term actually shows up in the case of the CF of two Wilson
lines.} the coefficients of the terms $\cot\theta$ and $\theta\cot\theta$ are
constrained by requiring that $\mathcal{C}_E(\theta)$ is crossing symmetric.
The analytic continuation of (\ref{Corr2E}) is
\begin{equation}\label{Corr2M}
\mathcal{C}_M(\chi)=\exp\bigg[i\left(\frac{K_1}{\sinh\chi}+K_2 \frac{\pi}{2}
\coth\chi + K_3 \cosh\chi\coth\chi\right)-\chi K_2\coth\chi\bigg]-1.
\end{equation}
The unitarity condition \eqref{unitcorrKM} becomes, in this case, $K_2\geq0$,
which is satisfied by the best--fit parameter within the errors
(see Table \ref{tab:corr123}).
The best--fit functions are plotted in Fig. \ref{corr2fit}.
After the best fit over the distances with an exponential function, one finds
for the leading--term coefficient $K_3$ a mass--scale
$\mu=3.79(1.46)\,\text{GeV}$. Thus, by
virtue of Eq. (\ref{sigmatotlead}) (with $n=1$, as implied by this
parameterization), this correlator leads to the following asymptotic total
cross section:
\begin{equation}\label{sigmatotcorr2}
\sigma^{(hh)}_{\text{tot}} \sim B \log^2 s,
\qquad \text{with:}\qquad B = 0.170^{+0.277}_{-0.081} \,\,\, \text{mb},
\end{equation}
that is again compatible, within the large errors, with the experimental result.

\newsubsubsection{Correlator 3}

The last parameterization that we are going to propose is:
\begin{equation}\label{Corr3E}
\mathcal{C}_E(\theta)=\exp\bigg[\frac{K_1}{\sin\theta} +
K_2 \left(\frac{\pi}{2}-\theta\right)^3 \cos\theta\bigg]-1.
\end{equation}
The first term is the usual $1/\sin\theta $, while the second one is less
\virg{familiar}, in the sense that it is not present in the analytical models
known in the literature: but is a fact that, using this parameterization,
the best fit is extremely good (see Table 1), even if it has only two
parameters. The Minkowskian version of the correlator (\ref{Corr3E}) is
\begin{equation}\label{Corr3M}
\mathcal{C}_M(\chi)=\exp\bigg[i\Big(\frac{K_1}{\sinh\chi} +
K_2 \cosh\chi\big(\frac{3}{4}\pi^2\chi-\chi^3\big)\Big) +
K_2\cosh\chi\big(\frac{\pi^3}{8}-\frac{3}{2}\pi\chi^2\big)\bigg]-1.
\end{equation}
The unitarity condition (\ref{unitcorrKM}) reduces (in the large--$\chi$ limit)
to $K_2\geq0$, that is fully satisfied by the best--fit parameter shown in
Table \ref{tab:corr123}.
The best--fit functions are plotted in Fig. \ref{corr3fit}.
As regards the total cross section, let us note that in this case the leading
term (for $\chi\to+\infty$) in the exponent in \eqref{Corr3M} is of the form
$\chi^3\,e^\chi$: so, we can find the asymptotic behavior of the total cross
section simply taking $f(\chi)=\chi^3\,e^\chi$ (i.e., with our notation,
$n=1$ and $p=3$) in the expression (\ref{intwithmandn}), that leads again
to the leading behavior reported in Eq. (\ref{sigmatotlead}) (with $n=1$).\\
After an exponential best fit over the distances, one finds that the
mass--scale of the leading--term coefficient $K_2$ is
$\mu=3.18(98)\,\text{GeV}$.
Thus, the asymptotic total cross section, derived from this correlator, reads:
\begin{equation}\label{sigmatotcorr3}
\sigma^{(hh)}_{\text{tot}} \sim B \log^2 s,
\qquad \text{with:}\qquad B = 0.245^{+0.263}_{-0.100} \,\,\, \text{mb}.
\end{equation}
The comparison with the experimental asymptotic coefficient is extremely good
and seems better than the previous ones, even if the errors are always very
large.

\newsection{Conclusions}

The nonperturbative approach to {\it soft} high--energy hadron--hadron
(dipole--dipole) scattering, based on the analytic continuation of
Wilson--loop CFs from Euclidean to Minkowskian theory,
makes possible the investigation of the problem of the asymptotic energy
dependence of hadron--hadron total cross sections from the point of view
of lattice QCD, by means of Monte Carlo numerical simulations.

In this paper we have performed a \emph{new} analysis of the data for the
Wilson--loop correlator, originally obtained in Refs. \cite{GM2008,GM2010}
by Monte Carlo simulations in Lattice Gauge Theory, and, in particular,
Section 4 has been focused on the search for a \emph{new}
parameterization of the (Euclidean) correlator that, in order:
$i)$ fits well the lattice data; $ii)$ satisfies (after analytic continuation)
the unitarity condition; and, most importantly, $iii)$ leads to a rising
behavior of total cross sections at high energy, in agreement with experimental
data. In particular, one is interested in the dependence of the correlation
function on the angle $\theta$ between the loops, since it is related, after
analytic continuation, to the energy dependence of the scattering amplitudes,
and also in its dependence on the impact--parameter distance. In Section 4 we
have shown that, making some reasonable assumptions about the angular
dependence and the impact--parameter dependence of the various terms in the
parameterization, our approach leads quite \virg{naturally} to total cross
sections rising asymptotically as $B \log^2 s$ (that is what experimental data
seem to suggest). Moreover, in our approach the coefficient $B$ turns out to be
\emph{universal}, i.e, the same for all hadronic scattering processes (as it
also seems to be suggested by experimental data), being related (see Eq.
(\ref{sigmatotlead})) to the mass--scale $\mu$ which sets the large
impact--parameter exponential behavior of the correlator: this type of behavior
is typical of a \emph{confining} theory, like QCD, and $\mu$ is expected to be
proportional to the lightest glueball mass $M_G$ or to the inverse of the
so--called \virg{\emph{vacuum correlation length}} $\lambda_{vac}$.
This is actually the main result of this paper.

Concerning the comparison between the numerical data obtained from the
best fits and the experimental value of $B$, the agreement is quite good
since the values are compatible within the large errors.
In Table \ref{tab:lambdavac} we report the mass--scale $\mu$ (and the
\emph{decay length} $\lambda=1/\mu$) derived from the parameterizations that
we have considered in the previous section, together with the predicted
universal coefficient $B=2\pi/\mu^2$.
\begin{table}
\centering
\begin{tabular}{|l|c|c|c|}
\hline
& $\mu$ (GeV) & $\lambda=\frac{1}{\mu}$ (fm) & $B=\frac{2\pi}{\mu^2}$ (mb) \\
\hline
Corr 1 & $4.64(2.38)$ & $0.042^{+0.045}_{-0.014}$ & $0.113^{+0.364}_{-0.037}$ \\
Corr 2 & $3.79(1.46)$ & $0.052^{+0.032}_{-0.014}$ & $0.170^{+0.277}_{-0.081}$ \\
Corr 3 & $3.18(98)$ & $0.062^{+0.028}_{-0.015}$ & $0.245^{+0.263}_{-0.100}$ \\
\hline
\end{tabular}
\caption{Comparison of the mass--scale $\mu$, the ``\emph{decay length}''
$\lambda=1/\mu$ and the coefficient $B=2\pi/\mu^2$ derived from our
parameterizations.}
\label{tab:lambdavac}
\end{table}
However, we want to remark the fact that the values that we have found have been
obtained from a limited set of \virg{short} (i.e., surely not asymptotic!)
distances, and so they must be taken only as an estimate. Of course, when
more lattice data (at larger distances) will be available, the relation
$B=2\pi/\mu^2 $ (derived from Eq. (\ref{sigmatotlead}) with $n=1$, as occurs
in our parameterizations of the correlator) may be confirmed or not.
Concerning the relation between $\mu$ and the inverse of the
\emph{vacuum correlation length} $\lambda_{vac}$ or the lightest glueball
mass $M_G$, at present a rigorous analytical determination in QCD is lacking
(apart from the result $\mu=2/\lambda_{vac}$ obtained in the SVM model) and
would be surely an important and helpful result. Using for $\mu$ an estimate
derived from the experimental value of $B$, i.e,
$\mu_{exp}=\sqrt{2\pi/B_{exp}}\simeq2.85$ GeV (as we have said above, our
numerical estimates for $\mu$ are compatible with $\mu_{exp} $ within the
large errors, as shown in Table \ref{tab:lambdavac}), one finds that
$\mu_{exp}\sim(3\div4)/\lambda_{vac}$ or $\mu_{exp}\sim2M_G$.
Of course, only further investigations (both numerical and analytical) can
confirm (or not) these results.
In this respect, we must also remark that the whole analytic
derivation of the result in Eq. (\ref{sigmatotlead}) is, of course, intended
to be performed in \emph{full} QCD (including dynamical quarks), while the
subsequent numerical analysis has been performed using the lattice data for the
Wilson--loop correlation function which are available at the moment, and which
were obtained in \emph{quenched} (i.e., \emph{pure--gauge}) QCD. However, we
expect that the mass--scale $\mu$, which enters Eq. (\ref{sigmatotlead}), is
essentially \emph{gluonic}, being related in some way (as we have said) to the
\emph{vacuum correlation length} $\lambda_{vac}$ or to the lightest glueball
mass $M_G$, and so it should not dramatically change when including dynamical
quark effects (hopefully, in near--future \emph{full}--QCD lattice computations
of the Wilson--loop correlator).
Pushing this ``speculation'' a little bit further, we indeed expect,
just on the basis of the experience with the \emph{vacuum correlation length}
$\lambda_{vac}$ (which increases from 0.22 fm in \emph{quenched} QCD up to
about 0.30 fm in \emph{full} QCD), that the inclusion of dynamical
quark effects should improve the agreement between the theoretical
determination of $\mu$ (i.e., of the \emph{decay length} $\lambda = 1/\mu$,
and of the parameter $B = 2\pi/\mu^2$: see the results in Table
\ref{tab:lambdavac}) and its experimental value $\mu_{exp} \simeq 2.85$ GeV
(corresponding to $\lambda_{exp} \simeq 0.07$ fm and $B_{exp} \simeq 0.3$ mb),
since $\lambda$ is expected to increase a little bit, so that $\mu$
should decrease and the parameter $B$ should increase, moving towards the
experimental value.
Of course, it would be desirable to have lattice results in full QCD $\dots$

Finally, let us observe that the functional integral approach turns out to be
fundamental for achieving this result: in fact, the investigation of
hadron--hadron elastic scattering in the \emph{soft} regime is mainly founded
on the \emph{elementary} loop--loop CF, which is then folded
with some proper wave functions for the specific hadrons involved in the
scattering process. Therefore, it is \virg{natural} to expect that a universal
behavior of the hadronic total cross sections at high energy may be originated
by the loop--loop correlator itself: and actually it has been so.
Of course, strictly speaking, our approach, based on the loop--loop
CF, and the corresponding conclusion about the universality
of $B$, only applies to {\it meson--meson} scattering: in this sense, we can
consider as a real {\it prediction} the fact that the value of $B$ that we have
found from our analysis is consistent (within the errors) with the experimental
value $B_{exp}$, which has been found considering {\it baryon--baryon}
(mainly, $pp$ and $p\bar{p}$) and {\it meson--baryon} scattering.
However, as briefly recalled at the beginning of Section 2, also
for the treatment of baryons a similar, but more involved,
picture can be adopted, using a genuine three--body configuration or,
alternatively and even more simply, a quark--diquark configuration
\cite{DFK,Nachtmann97,BN,Dosch,LLCM1}.
In particular, adopting a quark--diquark configuration for baryons,
we can directly extend our approach (based on the loop--loop correlator)
and the corresponding results to include also the case of {\it baryon--baryon}
and {\it meson--baryon} scattering.
This is probably enough to yield the {\it Pomeron} ($C$--even) contribution
(going as $B \log^2 s$) to hadron--hadron scattering, but not enough to
consider also possible {\it Odderon} ($C$--odd) contributions, which are
sub--leading in the high--energy limit, due to the {\it Pomeranchuk theorem}.
In fact, as already noticed above (in Section 4.3), these $C$--odd
contributions are averaged to zero in {\it dipole--dipole} scattering,
since in this case the relevant CF $C_E^{ave}$ is
automatically {\it crossing--symmetric}, and so they are probably visible
only adopting a genuine three--body configuration for baryons.

\section*{\large\bf Acknowledgements}
E.M. wants to thank Dr. Kazunori Itakura for useful discussions during the
Symposium of the YIPQS International Workshop ``High Energy Strong Interactions
2010'' at the Yukawa Institute for Theoretical Physics of Kyoto University,
which have stimulated this work.
M.G. has been supported by MICINN under the CPAN project CSD2007-00042
from the Consolider--Ingenio2010 program, and under the grant FPA2009-09638.

\newpage

%{\renewcommand{\Large}{\normalsize}
{\renewcommand{\Large}{\large}

}

\newpage

\pagestyle{empty}
\noindent
\begin{center}
%{\large\bf FIGURE CAPTIONS}
{\large\bf Figure captions}
\end{center}
\vskip 0.5 cm
\noindent

\begin{itemize}
\item[\bf Fig. 1]{The space--time configuration of the two Wilson loops
${\cal W}_1$ and ${\cal W}_2$ entering the expression for the dipole--dipole
elastic scattering amplitude in the high--energy limit.}
\item[\bf Fig. 2] {Comparison of lattice data for the averaged correlator
to best fits with the parameterization \eqref{Corr1E} (Correlator 1).}
\item[\bf Fig. 3] {Comparison of lattice data for the averaged correlator
to best fits with the parameterization \eqref{Corr2E} (Correlator 2).}
\item[\bf Fig. 4] {Comparison of lattice data for the averaged correlator
to best fits with the parameterization \eqref{Corr3E} (Correlator 3).}
\end{itemize}

%\cftpagenumbersoff{figure}
%\listoffigures

\newpage

\pagestyle{empty}

\begin{figure}[htb]
\includegraphics{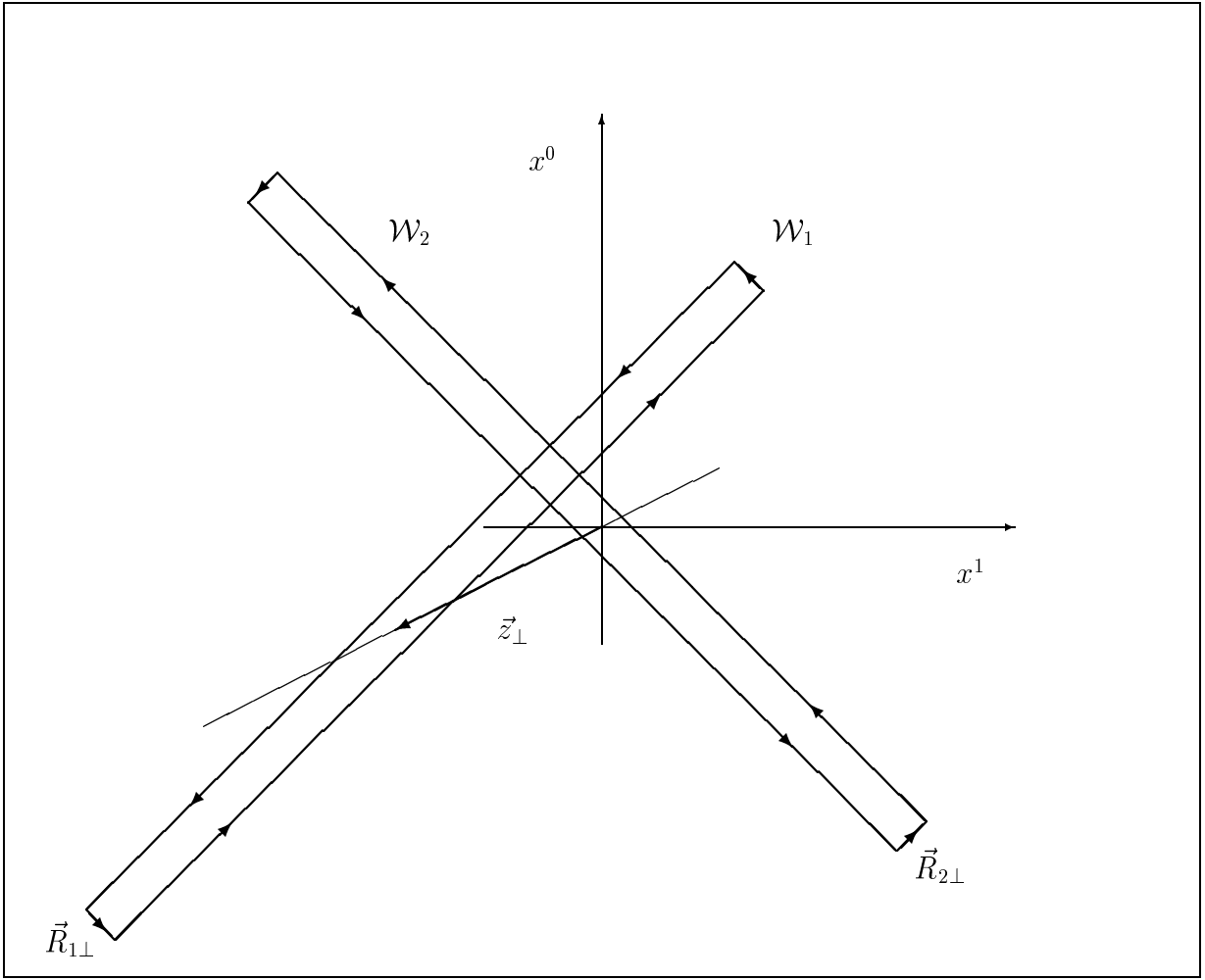}
\vspace{1.5cm}
\caption{The space--time configuration of the two Wilson loops ${\cal W}_1$ and
${\cal W}_2$ entering the expression for the dipole--dipole elastic
scattering amplitude in the high--energy limit.}
\label{fig:1}
\end{figure}

\clearpage

\begin{figure}
\centering
\resizebox{\textwidth}{!}{\includegraphics{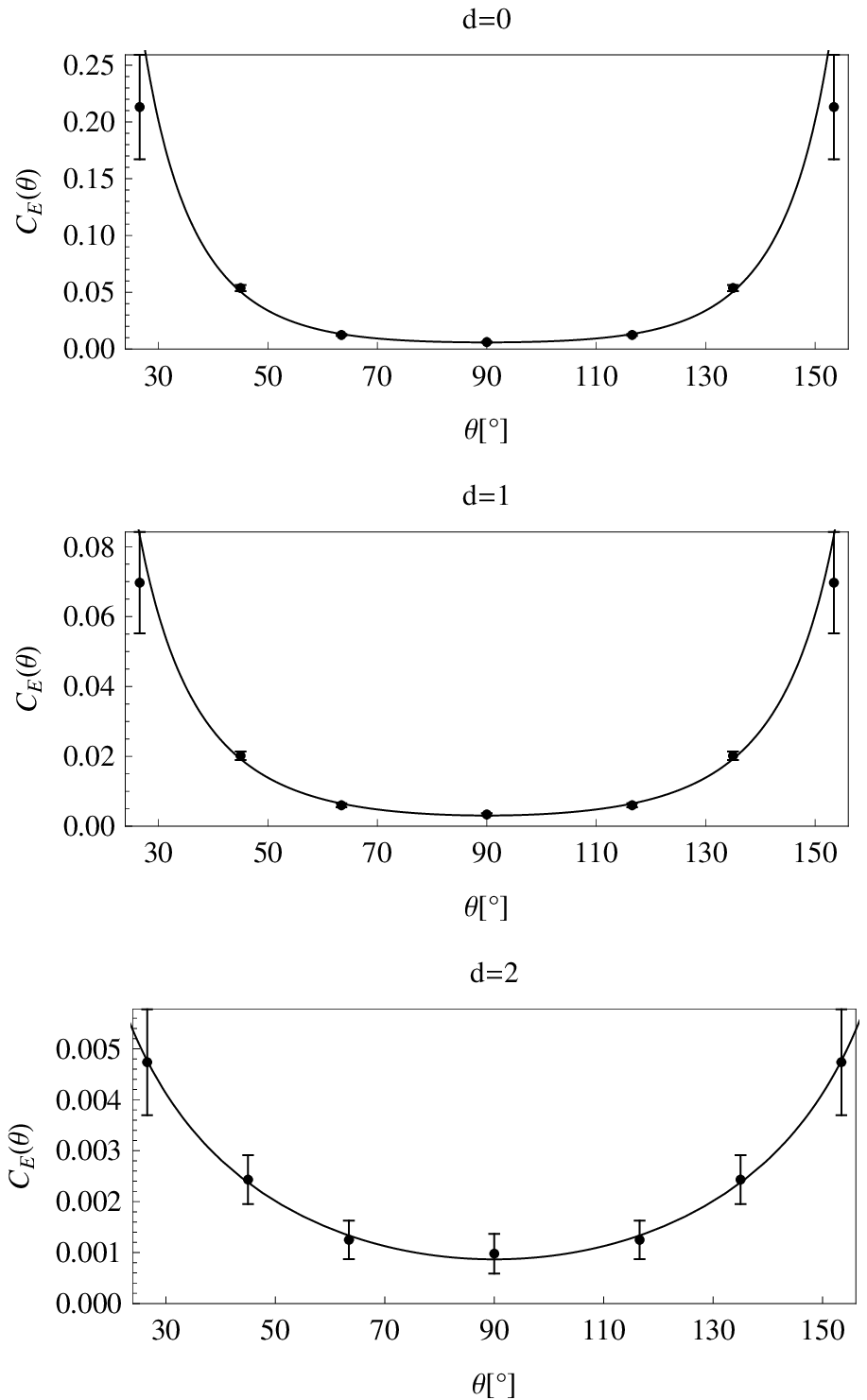}}
\vspace{1cm}
\caption{Comparison of lattice data for the averaged correlator to best fits
with the parameterization \eqref{Corr1E} (Correlator 1).}\label{corr1fit}
\end{figure}

\clearpage

\begin{figure}
\centering
\resizebox{\textwidth}{!}{\includegraphics{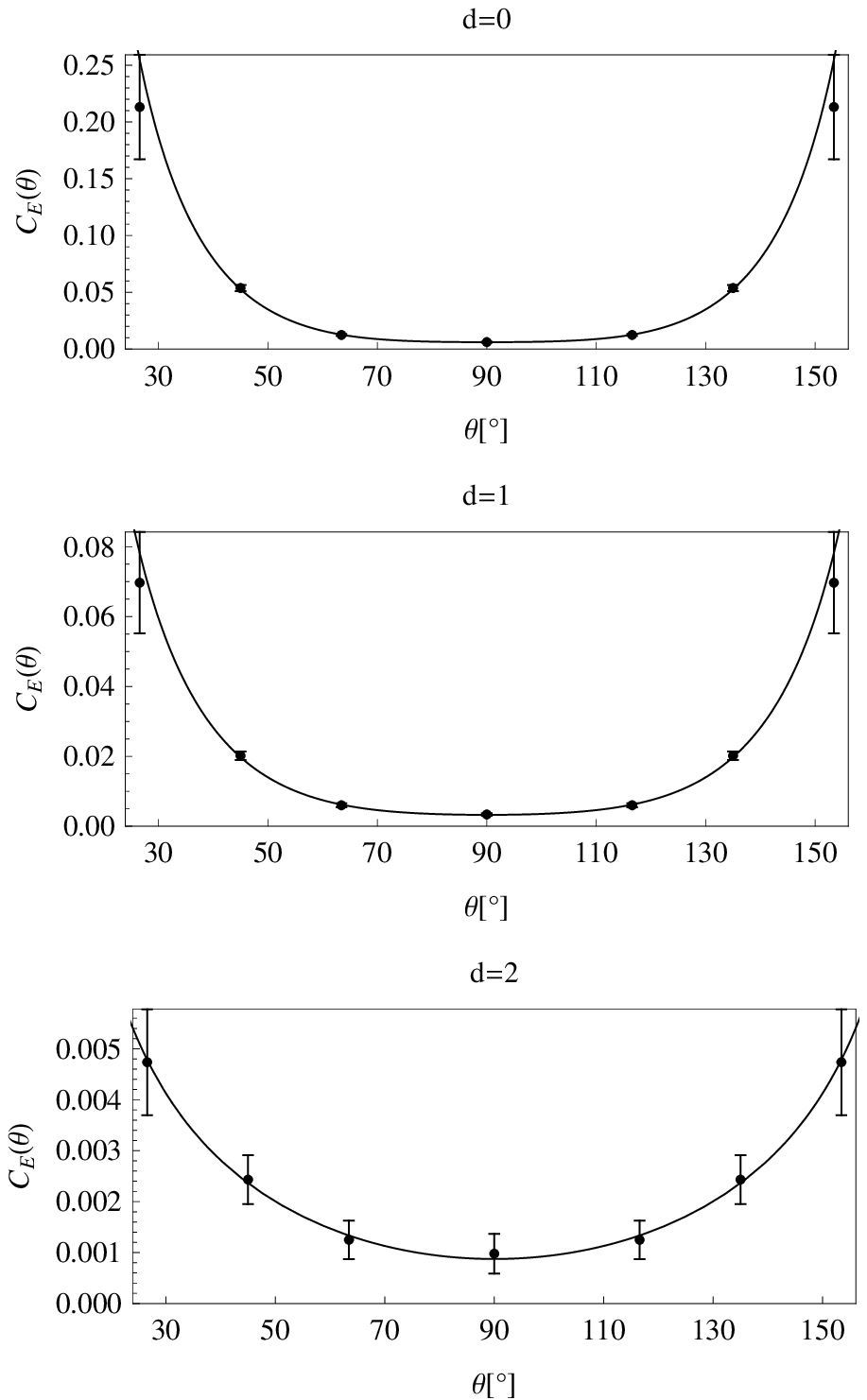}}
\vspace{1cm}
\caption{Comparison of lattice data for the averaged correlator to best fits
with the parameterization \eqref{Corr2E} (Correlator 2).}\label{corr2fit}
\end{figure}

\clearpage

\begin{figure}
\centering
\resizebox{\textwidth}{!}{\includegraphics{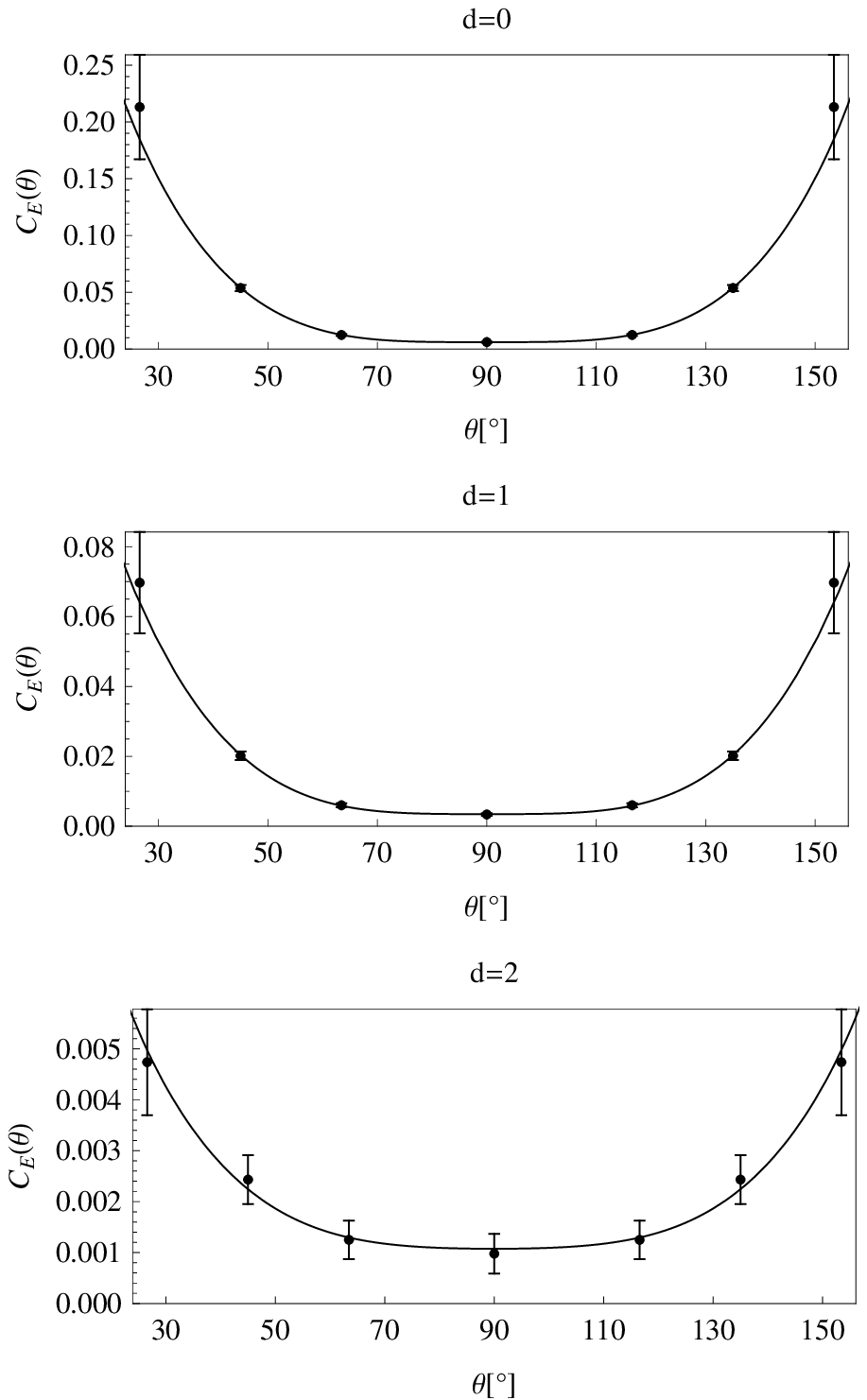}}
\vspace{1cm}
\caption{Comparison of lattice data for the averaged correlator to best fits
with the parameterization \eqref{Corr3E} (Correlator 3).}\label{corr3fit}
\end{figure}

\end{document}